


\font\titlefont = cmr10 scaled\magstep 4
\font\myfont    = cmr10 scaled\magstep 2
\font\sectionfont = cmr10
\font\authorfont = cmr12
\font\littlefont = cmr5
\font\eightrm = cmr8

\def\ss{\scriptstyle}
\def\sss{\scriptscriptstyle}

\newcount\tcflag
\tcflag = 0  

\ifnum\tcflag = 0 \magnification = 1200 \fi  

\global\baselineskip = 1.2\baselineskip
\global\parskip = 4pt plus 0.3pt
\global\abovedisplayskip = 18pt plus3pt minus9pt
\global\belowdisplayskip = 18pt plus3pt minus9pt
\global\abovedisplayshortskip = 6pt plus3pt
\global\belowdisplayshortskip = 6pt plus3pt


\def\endignore{}
\def\ignore #1\endignore{}

\newcount\dflag
\dflag = 0


\def\monthname{\ifcase\month
\or January \or February \or March \or April \or May \or June%
\or July \or August \or September \or October \or November %
\or December
\fi}

\newcount\dummy
\newcount\minute  
\newcount\hour
\newcount\localtime
\newcount\localday
\localtime = \time
\localday = \day

\def\advanceclock#1#2{ 
\dummy = #1
\multiply\dummy by 60
\advance\dummy by #2
\advance\localtime by \dummy
\ifnum\localtime > 1440 
\advance\localtime by -1440
\advance\localday by 1
\fi}

\def\settime{{\dummy = \localtime%
\divide\dummy by 60%
\hour = \dummy
\minute = \localtime%
\multiply\dummy by 60%
\advance\minute by -\dummy
\ifnum\minute < 10
\xdef\spacer{0} 
\else \xdef\spacer{}
\fi %
\ifnum\hour < 12
\xdef\ampm{a.m.} 
\else
\xdef\ampm{p.m.} 
\advance\hour by -12 %
\fi %
\ifnum\hour = 0 \hour = 12 \fi
\xdef\timestring{\number\hour : \spacer \number\minute%
\thinspace \ampm}}}



\def\endtitle{}
\def\title#1\endtitle{\vskip.5in\titlefont
\global\baselineskip = 2\baselineskip
#1\vskip.4in
\baselineskip = 0.5\baselineskip\rm}

\def\endauthors{}
\def\authors#1\endauthors{#1}

\def\endabstract{}
\def\abstract#1\endabstract{\vskip .3in%
\centerline{\sectionfont\bf Abstract}%
\vskip .1in
\noindent#1}

\def\nopageonenumber{\footline={\ifnum\pageno<2\hfil\else
\hss\tenrm\folio\hss\fi}}  

\newcount\nsection
\newcount\nsubsection

\def\section#1{\global\advance\nsection by 1
\nsubsection=0
\bigskip\noindent\centerline{\sectionfont \bf \number\nsection.\ #1}
\bigskip\rm\nobreak}

\def\subsection#1{\global\advance\nsubsection by 1
\bigskip\noindent\sectionfont \sl \number\nsection.\number\nsubsection)\
#1\bigskip\rm\nobreak}

\def\topic#1{{\medskip\noindent $\bullet$ \it #1:}}

\def\appendix#1#2{\bigskip\noindent%
\centerline{\sectionfont \bf Appendix #1.\ #2}
\bigskip\rm\nobreak}


\newcount\nref
\global\nref = 1

\def\therefs{}


\def\ref#1#2{\xdef #1{[\number\nref]}
\ifnum\nref = 1\global\xdef\therefs{\item{[\number\nref]} #2\ }
\else
\global\xdef\oldrefs{\therefs}
\global\xdef\therefs{\oldrefs\vskip.1in\item{[\number\nref]} #2\ }%
\fi%
\global\advance\nref by 1
}

\def\listrefs{\vfill\eject\section{References}\therefs}


\newcount\nfoot
\global\nfoot = 1

\def\foot#1#2{\xdef #1{(\number\nfoot)}
\footnote{${}^{\number\nfoot}$}{\eightrm #2}
\global\advance\nfoot by 1
}


\newcount\nfig
\global\nfig = 1
\def\thefigs{} 

\def\figure#1#2{\xdef #1{(\number\nfig)}
\ifnum\nfig = 1\global\xdef\thefigs{\item{(\number\nfig)} #2\ }
\else
\global\xdef\oldfigs{\thefigs}
\global\xdef\thefigs{\oldfigs\vskip.1in\item{(\number\nfig)} #2\ }%
\fi%
\global\advance\nfig by 1 } 

\def\figurecaptions{\vfill\eject\section{Figure Captions}\thefigs}

\def\fig#1{\xdef #1{(\number\nfig)}
\global\advance\nfig by 1 } 


\newcount\cflag
\newcount\nequation
\global\nequation = 1
\def\eqlabel{(1)}

\def\nexteqno{\ifnum\cflag = 0
\global\advance\nequation by 1
\fi
\global\cflag = 0
\xdef\eqlabel{(\number\nequation)}}

\def\lasteqno{\global\advance\nequation by -1
\xdef\eqlabel{(\number\nequation)}}

\def\label#1{\xdef #1{(\number\nequation)}
\ifnum\dflag = 1
{\escapechar = -1
\xdef\draftname{\littlefont\string#1}}
\fi}

\def\clabel#1#2{\xdef\eqlabel{(\number\nequation #2)}
\global\cflag = 1
\xdef #1{\eqlabel}
\ifnum\dflag = 1
{\escapechar = -1
\xdef\draftname{\string#1}}
\fi}

\def\cclabel#1#2{\xdef\eqlabel{#2)}
\global\cflag = 1
\xdef #1{\eqlabel}
\ifnum\dflag = 1
{\escapechar = -1
\xdef\draftname{\string#1}}
\fi}


\def\eeq{}

\def\eqnn #1\eeq{$$ #1 $$}

\def\eq #1\eeq{
\ifnum\dflag = 0
{\xdef\draftname{\ }}
\fi 
$$ #1
\eqno{\eqlabel \rlap{\ \draftname}} $$
\nexteqno}







\def\eqa #1\eeq{
\ifnum\dflag = 0
{\xdef\draftname{\ }}
\fi 
$$ \eqalignno{ #1 } $$
\global\cflag = 0}


\def\ie{{\it i.e.\/}}
\def\eg{{\it e.g.\/}}
\def\etc{{\it etc.\/}}
\def\etal{{\it et.al.\/}}


\def\mpla#1#2#3{{\it Mod.\ Phys.\ Lett.} {\bf A#1}, (19#2) #3}

\def\plb#1#2#3{{\it Phys.\ Lett.} {\bf #1B} (19#2) #3}

\def\prd#1#2#3{{\it Phys.\ Rev.} {\bf D#1} (19#2) #3}

\def\prl#1#2#3{{\it Phys.\ Rev.\ Lett.} {\bf #1} (19#2) #3}


\global\nulldelimiterspace = 0pt



\def\frac#1#2{{{#1} \over {#2}}\,}  



\def\Dsl{\hbox{/\kern-.6700em\it D}} 
\def\dsl{\hbox{/\kern-.5300em$\partial$}}
\def\pxpsl{\hbox{/\kern-.5600em$p$}}
\def\ssl{\hbox{/\kern-.5300em$s$}}
\def\epssl{\hbox{/\kern-.5100em$\epsilon$}}
\def\delsl{\hbox{/\kern-.6300em$\nabla$}}
\def\lxpsl{\hbox{/\kern-.4300em$l$}}
\def\elxpsl{\hbox{/\kern-.4500em$\ell$}}
\def\kxpsl{\hbox{/\kern-.5100em$k$}}
\def\qxpsl{\hbox{/\kern-.5000em$q$}}
\def\sla#1{\raise.15ex\hbox{$/$}\kern-.57em #1}



\def\roughly#1{\mathrel{\raise.3ex\hbox{$#1$\kern-.75em
  \lower1ex\hbox{$\sim$}}}}





\def\Scl{{\cal L}}


\def\ssl{{\sss L}}

\def\ssz{{\sss Z}}


\def\Re{{\rm Re\;}}






\def\MeV{{\rm \ MeV}}
\def\GeV{{\rm \ GeV}}

\overfullrule=0pt


\def\nc{{\rm nc}}

\def\rht{{\sss R}}
\def\lft{{\sss L}}
\def\sw{s_w}
\def\cw{c_w}

\def\Mz{M_{\sss Z}}

\def\mt{m_t}

\def\as{\alpha_s(M_{{\sss Z}})}

\def\leff{\Scl_{\rm eff}}

\def\gl{g_\lft}
\def\gr{g_\rht}
\def\hl{h_\lft}
\def\hr{h_\rht}

\def\zbb{Zb\overline{b}}

\def\ALR{A_{\sss LR}}
\def\AFB#1{A^0_{\sss FB}(#1)}

\def\Rb{$\hbox{\titlefont R}_{\hbox{\myfont b}}$}
\def\Rc{$\hbox{\titlefont R}_{\hbox{\myfont c}}$}

\voffset0.5in
\line{\hfill McGill/95-64, NEIP-95-014}

\vskip .6in
\centerline{\titlefont  \Rb , \Rc\ and New Physics:}
\centerline{\titlefont An Updated Model Independent Analysis}
\vskip .3in
\authors
\centerline{{\authorfont P.~Bamert} ${}^{a,b}$}
\vskip .3in
\centerline{\it ${}^a$ Physics Department, McGill University}
\centerline{\it 3600 University St., Montr\'eal, Qu\'ebec, CANADA, H3A
2T8.}
\vskip .05in
\centerline{\it ${}^b$ Institut de Physique, Universit\'e de Neuch\^atel}
\centerline{\it CH-2000 Neuch\^atel, Switzerland.}
\endauthors
\vskip .1in
\abstract
We analyze LEP and SLC data from the 1995 Summer Conferences as well as from
low energy neutral current experiments for signals of new physics. The reasons
for doing
this are twofold, first to explain the deviations from the standard
model observed in $R_b$ and $R_c$ and second to constrain non-standard
contributions
to couplings of the $Z^0$ boson to all fermions and to the oblique parameters.
We do so by comparing the data with the Standard Model as well as with a number
of
test hypotheses concerning the nature of the new physics. These include
non-standard
$Zb\bar{b}$-, $Zc\bar{c}$- and $Zs\bar{s}$-couplings as well as the couplings
of the
$Z^0$ to fermions of the entire first, second and third generations
and universal corrections to all up- and down-type quark couplings (as
can arise \eg\ in $Z'$ mixing models). 
We find that
non-standard $Zb\bar{b}$ couplings are both necessary and sufficient
to explain the data and in particular the $R_b$ anomaly. It is not possible to
explain
$R_b$, $R_c$ and a value of the strong coupling constant consistent with low
energy
determinations invoking only non-standard $Zb\bar{b}$- and
$Zc\bar{c}$-couplings.
To do so one has to have also new physics contributions to the $Zs\bar{s}$
or universal corrections to all $Zq\bar{q}$ couplings.
\endabstract


\vfill\eject
\section{Introduction}

\ref\lep{LEP electroweak working group and the LEP collaborations,''A
Combination of Preliminary LEP Electroweak Results and Constraints on the
Standard Model'', prepared form summer 1995 conference talks.}
\ref\slc{SLC Collaboration, as presented at CERN by C. Baltay (June 1995).}
\ref\qwcs{M. Noecker \etal, \prl{61}{88}{310}.}
\ref\ed{C.Y. Prescott \etal, \plb{84}{79}{524};
        P. Langacker and D. London, \prd{38}{88}{886}.}
\ref\nue{P. Vilain \etal, \plb{335}{94}{246}.}
\ref\bigfit{C.P. Burgess, S. Godfrey, H. K\"onig, D. London and I. Maksymyk,
          \prd{49}{94}{6115}.}
\ref\hagiwara{K. Hagiwara \etal, Talk presented at the fourteenth International
Workshop on Weak Interactions and Neutrinos in Seoul, hep-ph/9312231, (1993).}
\ref\langacker{P. Langacker, Invited Talk presented at SUSY-95, France,
hep-ph/9511207,
(1995).}
\ref\cdf{CDF Collaboration, F. Abe \etal , \prl{74}{95}{2626};
DO Collaboration, S. Abachi \etal , \prl{74}{95}{2632}.}
\ref\globalfit{P. Bamert, C.P. Burgess and I. Maksymyk, \plb{356}{95}{282}.}

The high precision measurements performed at LEP \lep\ and SLC \slc\ and
earlier on at lower energies \qwcs\ed\nue\bigfit\hagiwara\langacker , probing
the neutral
current interactions of the standard model (SM), have for a long time not
shown evidence for new physics. On the contrary, in the course of time their
agreement with the SM has become more and more precise, down to a level where
even weak radiative corrections had to be included into the theoretical
predictions for the observables. This sensitivity to lowest order weak
corrections eventually led to a prediction for the top mass well in agreement
with direct observations at CDF and D0 \cdf ; all in all a stunning
success and at the same time a frustrating experience for physicists who work
in a field where new discoveries are scarce.

Luckily, within the most recent period of LEP I data analysis an interesting
situation has emerged. There is for the first time a statistically significant
indication for physics beyond the standard model. The deviation, a surplus of
bottom quarks produced in $Z^0$ decays, has been there for some time but has
never
been large enough to warrant the effort of further investigation. This has
changed however with the inclusion of 1994 data. $R_b$, the width of the
$Z^0$-boson
to b quarks normalized to the total hadronic width, now lies some $3.7 \sigma$
above the SM prediction \lep .
This is not all the news however. Whereas $R_b$ increased, $R_c$, the width
to charm quarks (normalized in the same way),
decreased to be about $2.5\sigma$ below its SM value.
Although this deficit of $c$ quarks can still be viewed as a mere statistical
fluctuation it allows additional statements about the nature of possible new
physics when contrasted with $R_b$.

Given these exciting news people have begun to wonder at what kind of new
physics could explain these deviations. The aim of this letter is to help
them facilitate their tasks, by providing an analysis of the data in terms
of an effective lagrangian \bigfit\ that serves to parametrize the indirect
effects of heavy new physics. Using an effective lagrangian as a means of
characterizing physics beyond the SM has three main advantages: (i) it is
relatively
model independent, (ii) re-expresses the results of high precision electroweak
measurements in terms of quantities that are more straightforward to compute
within a given explicit model, thus serving as a kind of interface
between theory and experiments, and (iii) it approaches the problem with a
minimal set of assumptions about the nature of the new physics.

This letter also updates an earlier analysis \globalfit, which was based on
winter 1995 LEP data releases. Given the significant changes since, a fresh
and new analysis is not only justified, it also allows the drawing of new
conclusions.

This note is organized as follows: The next section reviews briefly the
effective lagrangian that has been derived in \bigfit\ and is used here.
In section three we discuss the observables we use in our fits, the results of
which are being presented in section four where we compare and analyze
different
hypotheses as to the nature of the new physics. Finally, most of the results
are concisely summarized in tables for quick reference.

\section{The effective lagrangian and parameter counting}

\ref\pestak{B. Holdom and J. Terning, \plb{247}{90}{88};
M.E. Peskin and T. Takeuchi, \prl{65}{90}{964}; \prd{46}{92}{381};
W.J. Marciano and J.L. Rosner, \prl{65}{90}{2963};
D.C. Kennedy and P. Langacker, \prl{65}{90}{2967}.}

We parametrize the indirect effects of heavy new physics in terms of the
effective lagrangian derived in \bigfit . In doing so we consider all potential
operators up to and including dimension four. These describe non-standard
contributions to the neutral current couplings, the charged current couplings
and
to the gauge boson self-energies\foot\efflagr{Such non-standard couplings can
emerge e.g.
when as yet undiscovered new particles give rise to additional radiative
corrections
to SM vertices or gauge boson self-energies.}.
Not including higher dimensional operators, such as four fermion contact
interactions,
is a shortcoming of this analysis. Such terms are however likely to be
suppressed, compared
to lower dimensional terms, by powers of the
typical scale of the new physics. Also, to be seen at LEP~I or SLC they would
have to compete
against $Z^0$-resonance amplitudes. This general approach allows us to compare
a large
variety of different new physics scenarios with the data, while at the same
time
invoking a minimal amount of theoretical prejudice.

We define the effective contributions to the neutral current couplings through
\eq\label\efflagr
\leff^{\nc} =
-\; {e\over \sw \cw} \overline{\psi}_f \gamma^\mu \left[ \left( \gl^f + \delta
\hat{g}_\lft^f
\right) \gamma\lft +  \left(\gr^f  + \delta \hat{g}_\rht^f \right) \gamma\rht
\right]
\psi_f Z_\mu ,
\eeq
where the SM couplings $\gl^f$ and $\gr^f$ are defined in terms of the quantum
numbers of the
corresponding fermion ($f$): $\gl^f = I_3^f - Q^f\sw^2$ and $\gr^f =
-Q^f\sw^2$, with $I_3$
being the weak isospin, $Q$ the electric charge and $\sw$ the sine of the weak
mixing angle. We consider here only flavor diagonal couplings, flavor violating
ones
being already stringently constrained (see \eg\ \bigfit ).

Similarly one can write down an effective charged current interaction, with
$\delta
h^{ff'}_L$ and $\delta h^{ff'}_R$ denoting the nonstandard contributions to the
couplings,
which are normalized such that the SM coupling of the $W$ boson to leptons
is $h^{ff'}_L = \delta_{ff'}$.

New physics contributions to gauge boson self energies finally can be
conveniently
parametrized in terms of the well known Peskin-Takeuchi parameters $S$, $T$ and
$U$
\pestak .

Using this parametrization, observables can now easily be expressed in terms of
their
SM expression and a new physics correction, linearized in $\delta \hat{g}_{\lft
,\rht}^f$,
$\delta h_{\lft ,\rht}^{ff'}$ and $S$, $T$ and $U$ \bigfit .

In this letter we will focus only on neutral current scattering processes,
since that is
where the deviations from the SM have been reported. Inclusion of charged
current
observables is straightforward, but does not add much to resolve the $R_b$
anomaly. At first glance it appears therefore that we need only consider the
nonstandard neutral current couplings $\delta \hat{g}_{\lft ,\rht}^f$ as well
as $S$
and $T$ \foot\upar{$\ss U$, representing a correction to the $\ss W$ self
energy, is not
being constrained by neutral current processes.}. Two specific combinations of
the
nonstandard charged current couplings
however do enter the analysis , through contributing to some of
the input quantities that are needed to derive the observables SM predictions.
In the case
of $Z^0$-pole data it is the Fermi coupling $G_F$, measured in muon decay, that
is used
together with the $Z^0$ mass $\Mz$ and the fine structure constant $\alpha$ as
an input.
In low energy neutrino-electron and neutrino-nucleon scattering the
$\delta h_{\lft ,\rht}^{ff'}$ enter because these processes are being
normalized to their charged current counterparts.

At the end of the day this means that we have to consider the following two
combinations of
non-standard charged current couplings when analyzing neutral current data
\bigfit :
\eq\label\deltas
\eqalign{\Delta &\equiv \Delta_e+\Delta_\mu\cr
\Delta^{\rm LE} &\equiv \Delta_e-\Delta_\mu+{\Re(\delta\hr^{ud})\over
|V_{ud}|}\cr}
,\quad\hbox{where
}\quad\Delta_f\equiv\sqrt{\sum_i|\delta_{if}+\delta\hl^{\nu_if}|^2}-1
\eeq
Here $V_{ud}$ is a CKM matrix element.
When linearized in the non-standard charged current couplings the $\Delta_f$
become
$\Delta_f\approx\Re(\delta\hl^{\nu_ff})$.

Using values for $e$ and $\sw$ in eq.\efflagr\ that are inferred from the input
observables
$\alpha$ and $G_F$ {\it assuming only SM physics} redefines parameters $\delta
\hat{g}_{\lft,\rht}^f$ such as to incorporate $S$, $T$ and $\Delta$ \bigfit :
\eq
\delta \hat{g}_{\lft,\rht}^f = {g_{\lft,\rht}\over 2}(\alpha T-\Delta) -
{Q^f\over (\cw^2-\sw^2)}
\left[\alpha S - \sw^2\cw^2(\alpha T-\Delta )\right] + \delta g_{\lft,\rht}^f
\eeq
where the $\delta g_{\lft,\rht}^f$ are the actual vertex corrections that come
about when
integrating out the new physics, so that the $\delta\hat{g}_{\lft,\rht}^f$
already account
for oblique corrections. From this it becomes clear that in any neutral current
process, as
long as $G_F$ is used as an input, only one linear combination of $T$ and
$\Delta$,
namely $T - \Delta /\alpha$, can be constrained whereas nothing can be said
about the other
one. So when we speak of $T$ in the analysis following below, we actually
mean this combination. Similarly, because non-standard couplings to strange
quarks enter the
neutral current observables only via their contribution to the total hadronic
width of the
$Z^0$ boson,
$\Gamma_{\rm had}$, they cannot be individually constrained. Instead we can
only put limits
on the following combination:
\eq\label\strange
\delta_s = \gl^s\delta\gl^s+\gr^s\delta\gr^s
\eeq
This leaves us with a set of 18 new-physics parameters: $\{\delta
g_{\lft,\rht}^f,\delta_s,S,T,
\Delta^{\rm LE}\}$ with $f=e,\mu,\tau,u,d,c,b$. When analyzing the data we will
subject
these parameters to various constraints, reflecting different hypotheses
about how the new physics dominantly couples.

If one were to consider, for example,
new physics that only couples to the third generation at tree level,
it's contributions to first and second
generation $Z^0f\bar{f}$ couplings would be highly suppressed, so that one can
restrict
the analysis to $\delta g_{\lft,\rht}^b,g_{\lft,\rht}^\tau ,S$ and $T$ while
constraining the other parameters to be zero.

\section{The Observables}

Before embarking to analyze the data in terms of the new physcis parameters
described above,
we first have a look at the various observables we consider. By cleverly
looking at how
these observables are linked it is possible to derive some conclusions about
the nature
of the new physics without having to refer to the power of an effective
lagrangian. These
statements will be quantified in the next section, where we actually discuss
the fits, and in
the tables.

Table~I shows all the observables considered in the present analysis. It is
divided into three
sections the uppermost of which lists the $Z^0$-pole observables of LEP and SLC
\lep\slc.
The next part of table~I displays the low-energy neutral current observables.
They stem from
atomic parity violation measurements in Cesium \qwcs, weak electromagnetic
interference
effects in $e$-$D$ experiments \ed, neutrino-electron \nue\ and deep inelastic
neutrino-nucleon
scattering \bigfit. Correlations between all these observables \lep\hagiwara\
are not displayed
in this table, but have been taken into account. The lowest segment of table~I
finally shows
the remaining input parameters. $\sin^2\theta_W$ as quoted here represents
Fermis constant
$G_F$ from which it has been derived. In this whole analysis the Higgs mass has
been fixed
to a fiducial value of $M_{\rm Higgs}=300\GeV$.

The low-energy observables of table~I only constrain non-standard contributions
to first
generation couplings and therefore serve no purpose in fits where these
couplings are
constrained to be zero. They have therefore not been included, with two
exceptions, to any
of the fits described later on so that the present analysis is mainly an
analysis of
$Z^0$-pole data.

{}From a first glance at table~I one
sees that, with four exceptions, all of the observables are in perfect
agreement with the SM.
This is especially true for the ones representing low energy experiments.
The four exceptions are
two asymmetry observables, $\AFB{\tau}$ and $\ALR^0$, as well as the two, by
know well known,
branching ratios to heavy quarks, $R_b$ and $R_c$.

Although $\AFB{\tau}$ and $\ALR^0$ differ from their SM values by $2.0\sigma$
and $2.4\sigma$ respectively it is hard to account for them using
only the indirect effects of new physics. This is easily seen through recalling
their
theoretical expressions, valid on resonance:
\eq\label\asym
\ALR^0={\cal A}_e,\quad\AFB{f} = {3\over 4}{\cal A}_e{\cal A}_f
\hbox{\ where \ }{\cal A}_f ={{g_L^f}^2-{g_R^f}^2
\over {g_L^f}^2 + {g_R^f}^2}
\eeq
So all the non standard couplings of our effective lagrangian can do is to
alter the values
of ${\cal A}_f$. Now $\ALR$ measures ${\cal A}_e$ as do two other observables,
${\cal A}_e(P_\tau)$ and $\AFB{e}$, which agree well with the SM. It is
therefore clear that this
discrepancy cannot be explained by heavy new physics. All this approach can do
is to alter the
value of ${\cal A}_e$ such as to interpolate between the two extremes and
thereby to reduce
the $\chi^2$ of the fit a bit. The same happens with $\AFB{\tau}$. Being
proportional to both
${\cal A}_e$ and ${\cal A}_\tau$ it calls for a deviation in one of these two
quantities.
${\cal A}_\tau$ as inferred from the $\tau$ polarization and angular
distribution of $\tau$'s, ${\cal A}_\tau (P_\tau )$, is however again in
excellent agreement with
the SM. This inability of the effective lagrangian approach pursued here to
fully explain the
deviations in the asymmetry observables argues for either a type of new physics
that is not
covered by this formalism, or it is an indication for not well understood
systematic effects in
the experiments.

\ref\basguys{B. Holdom, \plb{339}{94}{114};
J. Erler and P. Langacker, \prd{52}{95}{441};
G. Altarelli, R. Barbieri and F. Caravaglios, \plb{349}{95}{145}.}
\ref\shifmanI{M. Shifman, \mpla{10}{95}{605}.}
\ref\shifmanII{M. Shifman, preprint TPI-MINN-95/32-T, hep-ph/9511469 (1995).}

Turning now to the branching ratios, it is clear that any mechanism that solely
increases
$\Gamma (Z^0\rightarrow b\bar{b})$, and hence $R_b$, also increases the total
hadronic width,
which in turn increses the leptonic branching ratios $R_l = \Gamma_{\rm
had}/\Gamma_l$,
pushing them away from their SM values. As has been recognized by several
authors about a year
ago \basguys , one way of compensating for that increase is to lower the value
of the strong
coupling constant, on which $R_b$ does not depend\foot\rb{The QCD corrections
to the quarkian
decay widths $\ss\Gamma (Z^0\rightarrow q\bar{q})$ are universal. Any
dependence on the strong
coupling constant therefore cancels out in ratios such as $\ss
R_b=\Gamma_b/\Gamma{\rm had}$.}.
The value for $\alpha_s$ so obtained is
in agreement with low energy determinations. This in itself was a nice
surprise, because
as elaborated by Shifman \shifmanI\ there is a qualitative difference between
'low' values for
$\alpha_s\sim 0.112$ and high values $\alpha_s\sim 0.123$, such as derived \eg\
from SM fits
to LEP data. As argued in \shifmanI\ and \shifmanII\ this is due to the success
of QCD sum rules
and operator product expansions in low energy QCD, which require a low value of
the QCD scale
$\Lambda\sim 200\MeV$ (\ie\ a low value for $\alpha_s$ at the $Z^0$-scale)
and would fail for $\Lambda\sim 500\MeV$,
corresponding to a large $\alpha_s$. This argument can, and has been, turned
around by requiring
$\alpha_s$ to be small in fits of new physcis. For instance, including this
constraint, the
apparently low deviation of $R_b$ (around $2.4\sigma$) in Winter `95 data could
be translated
to a $3.5\sigma$ deviation \globalfit . Also, since explaining $R_c$
simultaneously with $R_b$
actually decreases $\Gamma_{\rm had}$, therefore demanding a very high value
for $\alpha_s$ to
compensate the damage done to the leptonic branching ratios, $R_c$ has often
been viewed as
a mere statistical fluctuation \shifmanII. One is of course free to do so,
since the costs
in $\chi^2$ when fitting for new physics are still bearable compared to those
one would have
to pay when not explaining $R_b$. As will be shown in the next section there is
however a
loophole which allows one to explain $R_c$ without having to refer to a high
value for $\alpha_s$.

There are many independent low-energy determinations of the strong coupling
constant
(see \eg\ \shifmanII ). In order to incorporate them in our fits we represent
them
through a 'pseudo observable'
\eq\label\asle
\alpha_s^{\rm LE}(M_Z) = 0.112\pm 0.003
\eeq
It should be cautioned however to take this value too literally. This is
because the
different determinations of $\alpha_s$ at low energies are subject to sometimes
very
different and partially unknown systematic errors \shifmanII. We merely use the
value
quoted above to {\it qualitatively} discriminate between different sets of new
physics.

\section{Fits and Discussion}

In this section we discuss the results of fitting various hypotheses, of how
possible
new physics might look like, to neutral current observables. To do so we use,
in general,
only $Z^0$-pole observables, the low energy observables listed in table I being
invoked only
to constrain first generation couplings. In fits where the non-standard
contributions
to these couplings are constrained to vanish, low-energy observables would only
improve
on the total $\chi^2/{\rm d.o.f.}$ therefore obscuring ''bad'' fits. On the
other hand
we always use the whole set of $Z^0$-pole observables even if some of them
might not
constrain the new physics parameters allowed to vary, because that is where the
deviations
are observed.

\topic{(1) The SM fit}
Table II displays the result of a standard model fit, where no new physics
parameters are
added, and only $\alpha_s$ and $\mt$ are allowed to vary. This fit serves
both as a test of the
reliability and as a point of reference for subsequent fits. The fact that four
observables
(and especially $R_b$) deviate by two or more standard deviations from their
respective SM
value translates here into a very bad confidence level of only $1.2\%$ for this
fit\foot\conlev{ By confidence level (CL) we mean the value of a $\ss\chi^2$
distribution with $\ss n-m$ degrees of freedom, where n is the number of
observables and m denotes the number of parameters that are allowed to float.
Under the assumption that the underlying theory is correct, the CL gives
the probability that if one were to redo all the experiments completely with the
same
analysis one would get a worse fit (i.e. a larger $\ss \chi^2/{\rm d.o.f.}$). A
low CL
therefore argues against the assumption that the underlying theory
is a correct description of nature.}.

\figure\bcoup
{A fit of the $\zbb$ couplings $\delta g^b_{\sss L,R}$
to $Z^0$-pole data from the 1995 Summer Conferences. This figure represents
the global fit of table III. The four solid lines respectively denote the 1--,
2--, 3-- and the 4--sigma error ellipsoids. The SM prediction lies at the
origin, $(0,0)$,
and is about $3.2 \sigma$ away from the center of the ellipses.}

\topic{(2) Non-standard b-couplings}
This fit is displayed in table~III as well as in figure~\bcoup . Before we
discuss it,
first a few words of explanation for table~III. It gives the results of a
global fit, where
both $\delta\gl^b$ and $\delta\gr^b$ are allowed to float and also shows how
fit values
of observables change in this case. Also, just by looking at the central values
and
error bars of $\gl^b$ and $\gr^b$ a deviation of about $3.2\sigma$ from $0$
(their SM value)
is not obvious. This is because these two parameters are heavily correlated
($0.83$) which is
reflected in a tilted error ellipsoid (see figure~\bcoup ). Because of this we
also give the
linear combinations of $\delta\gl^b,\delta\gr^b$ which are optimal in the sense
that they
diagonalize the variance matrix, \ie\ are uncorrelated, as well as their
central values and
error bars. Finally the results of individual fits to both $\delta\gl^b$ and
$\delta\gr^b$
are also given, mainly because many models that aim at explaining $R_b$ through
radiative corrections do so by changing either one coupling but not both.

With the dominant discrepancy from the SM being in $R_b$ it is not surprising
to see
that allowing for non-standard b-couplings improves the fit considerably. Upon
inspection of table~III one sees that it is now $R_c$ which lies farthest from
its fit value, deviating by $2.3\sigma$.
Certainly something that can be viewed as a statistical fluctuation and even be
expected, given the number of observables. This is
reflected in reasonable (albeit not excellent) confidence levels for these fits
in the
$15$-$20\%$ range. Non-standard b-couplings therefore are sufficient to resolve
the
LEP anomalies. Other non-standard couplings, such as to charm quarks, are not
necessary although
they can improve the fits as we will see below.

Another important point to notice is that it is impossible to distinguish
between left- and right
handed b-couplings, $\delta\gl^b$ and $\delta\gr^b$. This is obvious from
looking at
figure~\bcoup , follows from the fact that the two are strongly correlated and
can be seen through
comparing the confidence levels of the two individual fits in table~III (being
$15\%$ and $19\%$
for $\delta\gl^b$ and $\delta\gr^b$ respectively).

The third, and probably most important point is, as has been mentioned earlier,
that these fits
point to low values of the strong coupling constant $\alpha_s$, well in
agreement with
low-energy determinations (see eq.~\asle ). By adding one parameter (\eg\
$\delta\gl^b$) one
can thus resolve two discrepancies, namely $R_b$ and the difference in
determinations of
$\alpha_s$. How strong these two discrepancies are linked can also be seen from
the fact that,
when fixing $\alpha_s$ to the SM fit value ($0.123$) the fit C.L. drops to
$2.8\%$. Including
$\alpha_s^{\rm LE}$ as given in eq.~\asle\ on the other hand, doesn't change
the fit C.L. very much
($14\%$) but increases the deviation in the b-couplings from $3.2\sigma$ to
$3.8\sigma$. This
is due to the smaller error bars of $\alpha_s^{\rm LE}$ compared to table~III.

\topic{(3) Other non-standard couplings}
Assuming the new physics to show up in other than the b-couplings does not
work, making
nonstandard b-couplings not only sufficient, but also necessary, within the
current framework,
to explain the LEP anomalies. Fit confidence levels are \eg\ $1.2\%$ for an
oblique fit ($S,T$),
$0.4\%$ for tau-couplings ($\delta g_{\lft,\rht}^\tau$), \etc ,
the list could be continued. All these parameters are unable to explain the
$R_b$ anomaly.

It is amusing to notice that the combination $\delta\gl^c,\delta\gr^c$ could
explain
both $R_c$ and $R_b$ where it not for an unacceptably large value of the strong
coupling constant, which would be required to be some $4.4\sigma$ away from the
value quoted
in eq.~\asle . This is because the absolute value of the deviation of $R_c$ is
much larger than
in $R_b$, hence explaining $R_c$ reduces $\Gamma_{\rm had}$ enough to account
for $R_b$ as well.
Unfortunately one then has to repair the damage done to the leptonic branching
ratios by
increasing $\alpha_s$ to ridiculous values. Upon including $\alpha_s^{\rm LE}$
of eq.~\asle\ in the
fit the C.L. drops to $0.2\%$, excluding the new physics to be exclusively in
the c-couplings.

\figure\obl
{The results of the global fit to $\delta g_{\lft,\rht}^b,S$ and $T$ shown
in table IV, projected to the $\delta g_{\lft,\rht}^b$-plane (above) and the
$S,T$-plane
(below). The upper graph is similar to the one in figure \bcoup , whereas in
the lower figure
only the 1 and 2-sigma error ellipsoids are shown. In the latter figure the
origin,
representing the SM for a fiducial value of the top mass of $m_t=180\GeV$,
lies some $1.5\sigma$ from the fit values for $S$ and $T$.}

\figure\ccoup
{The results of a global fit to non-standard left-handed b- and c-couplings
similar to
the one in table V. The four curves again denote the 1-4 sigma error
ellipsoids.
Although having an excellent confidence level this fit smacks of a very high
value for
the strong coupling constant.}

\topic{(4) Combining $\delta g_{\lft,\rht}^b$ with other new physics
parameters}
Once it has become clear that b-couplings are both necessary and sufficient to
explain the
LEP anomalies one can go one step further and see what kind of constraints  can
be found
for other new physics parameters. The results of these fits are summarized in
tables IV through
X and figures~\obl\ and \ccoup .

Since many models that yield radiative corrections to the $Zb\bar{b}$ vertex
also contribute
through extra gauge-boson self-energies to the oblique parameters we give in
table IV and in
figure~\obl\ the results of an oblique fit. It is mainly the difference between
LEP and SLC
asymmetry measurements that drives $S$ in the negative direction.

It is not surprising that by adding non-standard charm couplings to $\delta
g_{\lft,\rht}^b$
explains both, the $R_b$ and $R_c$ anomalies. In fact the new fit values of
these
observables are now shifted to coincide with their experimental values.
Consequently the fit has
an excellent confidence level of $37\%$. It is displayed in table V and
figure~\ccoup .
Notice that the c-couplings are only weekly correlated in this fit ($0.09$)
putting the
deviation entirely into $\delta\gl^c$. However, again, since the absolute
deviation of $R_c$
is larger than the one of $R_b$, the hadronic width $\Gamma_{\rm had}$ is
reduced in this scheme,
lowering the leptonic branching ratios $R_l$.\foot\cvsb{The reduced total
hadronic width also
increases $\ss R_b$ which is reflected in the fact that in this fit the charm
and not the
bottom couplings display the strongest deviation from zero.}
Correcting for that requires a value for
$\alpha_s$ ($= 0.18\pm 0.035$) which lies some $1.9\sigma$ above low-energy
determinations.
Including $\alpha_s^{\rm LE}$ from eq.~\asle\ in the fit not only reduces the
CL to $19\%$,
\ie comparable to the b-coupling fits of table~III, but also is unable to
explain $R_c$ (it
merely reduces the pull of this observable from $-2.5$ to $-1.9$). In this case
the
two non-standard c-couplings become also become strongly correlated.

Luckily there is a loophole in the above reasoning. Since both $\delta
g_{\lft,\rht}^b$ and
$\delta g_{\lft,\rht}^c$ can compensate $R_b$ and $R_c$ for any shift
$\Gamma_{\rm had}$ might recieve from elsewhere, one is not fixed to invoking
only
$\alpha_s$ to correct the leptonic branching ratios\foot\contr{This is contrary
to
the case of the c-coupling only fit.}. Non-standard contributions to the
light quark couplings for example could reduce $\Gamma_{\rm had}$ such as to
reconcile
$R_l$ with their SM values. First generation quark couplings are
constrained by low-energy neutral current observables (displayed in table~I) as
is shown in a corresponding fit in table~VIII. 
They would therefore have to be modified simultaneously to adjust 
$\Gamma_{\rm had}$ in the desired way. A more viable option is non-standard
strange quark couplings. The result of a series of fits for a set of different
fiducial values for $\alpha_s$ is shown in table~VI. Of course this argument
can be turned
around, {\it if one where to ignore low-energy determinations of $\alpha_s$},
in the sense that
the pure b-coupling fits shown in table~III can be reconciled with a high value
of the strong
coupling constant, such as favoured by GUT's, upon including new physics
contributions to
strange quarks ($\delta_s$). A corresponding fit, with $\alpha_s$ fixed to
$0.123$ for example,
yields $\delta_s= -0.00200\pm 0.00077$ and has a confidence level of $16\%$.

It is also interesting to see what happens when one allows the quark couplings
to change in an universal way. Such a situation can arise \eg\ in models with 
and extra $Z'$ gauge boson. We are then led to constrain the quark couplings
in the follwing way:
\eq\label\universal
\eqalign{\delta u_L \equiv &\;\delta g^u_L = \delta g^c_L\cr
\delta d_L \equiv &\;\delta g^d_L = \delta g^s_L= \delta g^b_L}
\eeq
and similarily for the right-handed couplings.
Such a fit has a very good confidence level since
the light quark couplings, all receiving corrections simultaneously, 
can do the job $\delta_s$ did above in compensating the damage done by 
$R_c$ to $\Gamma_{\rm had}$. The results of a fit floating all four degrees
of freedom ($\delta u_{L,R},\delta d_{L,R}$) is shown in table VII. Notice
that varying only two of the above parameters 
(such as \eg\ only corrections to 
the left-handed couplings $\delta u_L,\delta d_L$) doesn't work, with
the exception of the combination ($\delta u_R,\delta d_L$), because then
the low-energy observables usually forbid corrections as large as needed to
explain $R_b$ and $R_c$ (in the case of \eg\ only left-handed couplings it is
$Q_W$(Cs) that 'counteracts' $R_b$ and $R_c$). If on the other hand one were 
to impose the additional constraint that corrections to the up-type couplings 
equal those to down-type couplings, \ie\ $\delta_L\equiv\delta u_L=
\delta d_L$ and $\delta_R\equiv\delta u_R = \delta d_R$, the data can be 
explained. A corresponding fit yields $\delta_L=-0.00414\pm 0.00087$ and
$\delta_R = 0.0073\pm 0.0023$ with a correlation of $-0.5$ and a confidence
level of $62\%$. In this case $R_b$ would be $1\sigma$ above and $R_c$ $1.5
\sigma$ below the experimental values. Interpreted 
in terms of an extra gauge boson that
couples equally to all up- and down-type quarks these values favour an
axial coupling. A purely vectorial coupling on the other hand (where 
$\delta_L=\delta_R$) could not explain the data.

To complete the analysis we also give the results for fits to first, second and
third generation
couplings. They are displayed in tables~VIII through X respectively.
In the first two cases $\delta\gl^b$ was also allowed to vary in order to get a
reasonable fit confidence level. In the fit to the first generation couplings,
the low-energy
observables shown in table~I have been used and a fiducial value for $\alpha_s$
of $0.112$ has
been applied. These tables serve mainly to give limits on new physics
parameters other than
the ones discussed earlier on in this letter.

\topic{(5) Sensitivity to the top mass}
The various new physics fits that float $\mt$ all yield values which are
in the region of the top mass found in the SM fit of table~II (for a Higgs
mass of $300 \GeV $). The top mass is therefore not very sensitive to the new
physics scenarios considered here and allowing it to vary does not improve the
fits significantly. Also the top masses obtained in the different scenarios
are compatible with direct determinations from CDF and D0 \cdf . Notice that
the top mass is in some fits only weakly constrained because the chosen
new physics parameters can affect the observables that are most sensitive to
the top. This is for instance the case in a fit where $\mt$ is floated
simultaneously with $T$ and $\delta g^b_L$ which can compensate for the 
dominant ($\propto \mt^2$) radiative corrections to the observables. In fits
like this $\mt$ has therefore been fixed to a 
fiducial value\foot\asfid{The same observation holds for fits in 
which $\ss\as$ has been fixed to fiducial values.}.  

\section{Conclusions}

We have analyzed $Z^0$-pole and low energy neutral current measurements for
signals
of new physics. We were able to do so in a very general and model independent
way by
using an effective lagrangian which serves to parametrize indirect
contributions
of new physics to \eg\ $Z^0$-fermion couplings or gauge boson self energies.
The following points concisely summarize our results:
\item{*}Non-standard contributions to the $Z^0b\bar{b}$ couplings ($\delta
g_{\lft,\rht}^b$)
       are both necessary and sufficient to explain the data and in particular
the $R_b$ anomaly.
\item{*}It is not possible to decide whether the new physcis contributes to
       the left- or to the right-handed $Z^0b\bar{b}$ coupling.
\item{*}Allowing for $\delta g_{\lft,\rht}^b$ alone points to a low value for
the strong
       coupling constant, well in agreement with low energy determinations.
\item{*}The best fit invokes non-standard bottom as well as charm couplings. It
explains both
       the $R_b$ and the $R_c$ anomalies. It yields however a value for
$\alpha_s$ that
       is about two standard deviations above low-energy determinations
($\alpha_s^{\rm LE}$).
\item{*}The same fit, when including $\alpha_s^{\rm LE}$, does not explain the
$R_c$ deviation.
\item{*}It is possible to reconcile $R_b$, $R_c$ and $\alpha_s^{\rm LE}$ by
including
       non-standard contributions to strange quark as well as to bottom and
charm quark couplings. Or by allowing for universal corrections to all up and
to all down type quark couplings such as can occur in models with an extra $Z'$
gauge boson.
\item{*}There is no conclusive evidence for the new physcis to contribute to
$Z^0$-couplings other
       than the ones to bottom quarks ($\delta g_{\lft,\rht}^b$).

\bigskip
\centerline{\bf Acknowledgements}
\bigskip

The author would like to thank Cliff Burgess, David London and Oscar Hernandez
for many helpful and enlightening discussions and Georges Azuelos for providing
an early version of \lep . This research was financially supported by
the NSERC of Canada, the FCAR du Qu\'ebec and the Swiss National Foundation.

\listrefs

\vfill\eject



$$\vbox{\tabskip=0pt \offinterlineskip
\halign to \hsize{#\tabskip 1em plus 2em minus .5em&\hfil#\hfil
&\hfil#\hfil &\hfil#\hfil &\hfil#\hfil&\tabskip 1em plus 2em minus .5em #\cr
\noalign{\hrule}\noalign{\smallskip}\noalign{\hrule}\noalign{\medskip}
& \hfil \hbox{Quantity} & \hfil \hbox{Experimental Value}&
\hfil \hbox{Standard Model Fit} &\hbox{Pull}&\cr
\noalign{\medskip}\noalign{\hrule}\noalign{\medskip}
& $\Mz$ (GeV)  & $91.1885 \pm 0.0022 $  & input & ---& \cr
& $\Gamma_\ssz$ (GeV)  & $ 2.4963 \pm 0.0032 $  & $2.4973$ &$ -0.3$&
\cr
& $\sigma^h_p $ (nb)  & $41.488 \pm 0.078$ & $41.45$ &$0.5$& \cr
& $R_e=\Gamma_{\rm had}/\Gamma_{e}$ & $20.797 \pm 0.058$
	& $20.773$ &$0.4$& \cr
& $R_\mu=\Gamma_{\rm had}/\Gamma_{\mu}$  & $20.796 \pm 0.043$
	& $20.773$ &$0.5$& \cr
& $R_\tau=\Gamma_{\rm had}/\Gamma_{\tau}$  & $20.813 \pm 0.061$
	& $20.821$ &$-0.1$& \cr
& $\AFB{e}$ & $ 0.0157\pm 0.0028$  & $0.0159$ &$-0.1$& \cr
\noalign{\smallskip}
& $\AFB{\mu}$ & $ 0.0163\pm 0.0016$  & $0.0159$ &$0.3$& \cr
\noalign{\smallskip}
& $\AFB{\tau}$ & $ 0.0206\pm 0.0023$  & $0.0159$ &$2.0$& \cr
\noalign{\smallskip}
& $A_\tau(P_\tau)$  & $0.1418 \pm 0.0075$  & $ 0.1455$ &$-0.5$& \cr
\noalign{\smallskip}
& $A_e(P_\tau)$  & $ 0.139 \pm 0.0089$  & $ 0.1455$ &$-0.7$& \cr
\noalign{\smallskip}
& $R_b=\Gamma_b/\Gamma_{\rm had}$  & $0.2219 \pm 0.0017$ & $ 0.2156 $ &$3.7$&
\cr
& $R_c=\Gamma_c/\Gamma_{\rm had}$  & $0.154 \pm 0.0074$ & $ 0.1724 $ &$-2.5$&
\cr
& $\AFB{b}$  & $0.0997 \pm 0.0031 $  & $0.102$ &$-0.7$& \cr
\noalign{\smallskip}
& $\AFB{c}$  & $0.0729 \pm  0.0058$  & $ 0.0728$ &$0.0$&\cr
\noalign{\smallskip}
& $\ALR^0$  & $0.1551 \pm 0.0040$  & $0.1455$ &$2.4$&\cr
\noalign{\medskip}\noalign{\hrule}
\noalign{\smallskip}
& $Q_W$ (Cs)  & $-71.04 \pm  1.81$  & $ -72.88$ &$1.0$&\cr
& $g^e_A$ ($\nu$-e)  & $-0.503 \pm  0.017$  & $ -0.507$ &$0.2$&\cr
& $g^e_V$ ($\nu$-e)  & $-0.035 \pm  0.017$  & $ -0.037$ &$0.1$&\cr
& $C_{1u}-{1\over2}C_{1d}$ (e-D) & $-0.47 \pm  0.13$  & $ -0.361$ &$-0.8$&\cr
\noalign{\smallskip}
& $C_{2u}-{1\over2}C_{2d}$ (e-D) & $0.33 \pm  0.62$  & $ -0.039$ &$0.6$&\cr
& $g_L^2$ ($\nu$ -N)  & $ 0.3003 \pm  0.0039$  & $ 0.3021$ &$-0.5$&\cr
\noalign{\smallskip}
& $g_R^2$ ($\nu$ -N)  & $ 0.0323 \pm  0.0033$  & $ 0.0302$ &$0.6$&\cr
\noalign{\medskip}\noalign{\hrule}
\noalign{\smallskip}
& $\alpha(M_Z)$ & $1/128.896$ & input & --- &\cr
& $\sin^2\theta_W(G_F)$& $0.2311$& input& --- &\cr
& $M_{\rm Higgs}$ (GeV) & 300 & fiducial & --- &\cr
\noalign{\medskip}\noalign{\hrule}\noalign{\smallskip}\noalign{\hrule}
}}$$
\centerline{{\bf TABLE I: Observables}}
\smallskip
\noindent {\eightrm The experimental values for neutral current observables
considered in
the present analysis. The column labelled 'Pull' gives the difference between
the
experimental and the SM fit value of an observable in units of a standard
deviation. The
upper section of the table lists the various $\ss Z^0$-pole observables,
whereas the ones in
the second section stem from low energy experiments, such as atomic parity
violation and
scattering experiments. At the bottom of the table one finds the values we use
for the various
input quantities. The Fermi coupling $\ss G_F$, advertised as an input in the
text, is
represented here by $\ss\sin^2\theta_W$, obtained from $\ss G_F$, $\ss\alpha$
and $\ss M_Z$. The Standard Model Fit values in the third column are given here
for a top mass of $\ss m_t=180$ GeV and a strong coupling constant 
$\ss\alpha_s(M_Z)=0.123$. 
}
%


$$\vbox{\tabskip=0pt \offinterlineskip
\halign to \hsize{\strut#& #\tabskip 1em plus 2em minus .5em&\hfil#\hfil
&\hfil#\hfil &\hfil#\hfil &#\tabskip=0pt\cr
\noalign{\hrule}\noalign{\smallskip}\noalign{\hrule}\noalign{\medskip}
&& \hfil \hbox{Parameter} & \hfil Fit Value\hfil &\cr
\noalign{\medskip}\noalign{\hrule}\noalign{\medskip}
&& $\as$ & $0.123 \pm 0.004$ &\cr
&& $\mt \hbox{(GeV)}$ & $181 \pm 6$ &\cr
\noalign{\medskip}\noalign{\hrule}\noalign{\medskip}
&& $\chi^2_{\rm min}/\hbox{d.o.f.}$ & 27.2/13 (1.2\% C.L.) &\cr
\noalign{\medskip}\noalign{\hrule}\noalign{\smallskip}\noalign{\hrule}
}}$$
\centerline{{\bf TABLE II: Standard Model}}
\smallskip
\noindent {\eightrm The SM parameters ($\ss \as$ and $\ss \mt$)
as determined by fitting to the $\ss Z^0$-pole observables of Table I. }
%
\vfill\break

$$\vbox{\tabskip=0pt \offinterlineskip
\halign to \hsize{\hfil#\hfil&\hfil#\hfil&\hfil#\hfil
&\hfil#\hfil &#\hfil&\hfil#\hfil\cr
\noalign{\hrule}\noalign{\smallskip}\noalign{\hrule}\noalign{\medskip}
 \hfil \hbox{Parameter} & \hfil \hbox{Global Fit}&&
\hfil \hbox{Individual Fit}&&\hbox{Individual Fit}\cr
\noalign{\medskip}\noalign{\hrule}\noalign{\medskip}
$\delta\gl^b$&$-0.0029\pm 0.0037 $& $(0.8)$&$-0.0067\pm 0.0021$& $(3.2)$&\cr
\noalign{\smallskip}
$\delta\gr^b$&$ 0.022 \pm 0.018  $& $(1.2)$&&&$0.034\pm 0.010$ $(3.3)$\cr
\noalign{\medskip}
$ 0.986\delta\gl^b-0.167\delta\gr^b$&$-0.0066\pm 0.0021 $& $(3.2)$&&&\cr
\noalign{\smallskip}
$ 0.167\delta\gl^b+0.986\delta\gr^b$&$ 0.022 \pm 0.019  $& $(1.2)$&&&\cr
\noalign{\medskip}
$\alpha_s({\ss M_Z})$&$ 0.101 \pm 0.007 $&&$ 0.103 \pm 0.007$&&$0.103\pm 0.007$
\cr
$\mt$ (GeV)&$ 188 \pm 7       $&&$ 185 \pm 6      $&&$190\pm 6$\cr
\noalign{\medskip}
$\chi^2/{\rm d.o.f.}$&$15.5/11$ ($16\%$ C.L.)&&$17.0/12$ ($15\%$ C.L.)&&
16.1/12 ($19\%$ C.L.)\cr
\noalign{\medskip}\noalign{\hrule}\noalign{\medskip}
 Observable & Experimental Value && New Fit Value && New Pull\cr
\noalign{\medskip}\noalign{\hrule}\noalign{\medskip}
 $\Gamma_\ssz$ (GeV) & $2.4963 \pm 0.0032 $ && $2.4992$ &&-0.9\cr
 $\sigma^h_p $ (nb)  & $41.488 \pm 0.078  $ && $41.461$ && 0.3\cr
 $R_e$               & $20.797 \pm 0.058  $ && $20.763$ && 0.6\cr
 $R_\mu$             & $20.796 \pm 0.043  $ && $20.763$ && 0.8\cr
 $R_\tau$            & $20.813 \pm 0.061  $ && $20.811$ && 0.0\cr
 $\AFB{e}$           & $0.0157 \pm 0.0028 $ && $0.0165$ &&-0.3\cr
\noalign{\smallskip}
 $\AFB{\mu}$         & $0.0163 \pm 0.0016 $ && $0.0165$ &&-0.1\cr
\noalign{\smallskip}
 $\AFB{\tau}$        & $0.0206 \pm 0.0023 $ && $0.0165$ && 1.8\cr
\noalign{\smallskip}
 $A_\tau(P_\tau)$    & $0.1418 \pm 0.0075 $ && $0.1481$ &&-0.8\cr
\noalign{\smallskip}
 $A_e(P_\tau)$       & $0.139  \pm 0.0089 $ && $0.1481$ &&-1.0\cr
\noalign{\smallskip}
 $R_b$               & $0.2219 \pm 0.0017 $ && $0.2207$ && 0.7\cr
 $R_c$               & $0.154  \pm 0.0074 $ && $0.1713$ &&-2.3\cr
 $\AFB{b}$           & $0.0997 \pm 0.0031 $ && $0.1005$ &&-0.2\cr
\noalign{\smallskip}
 $\AFB{c}$           & $0.0729 \pm 0.0058 $ && $0.0747$ && -0.3\cr
\noalign{\smallskip}
 $\ALR^0$            & $0.1551 \pm 0.0040 $ && $0.1481$ && 1.7\cr
\noalign{\medskip}\noalign{\hrule}\noalign{\smallskip}\noalign{\hrule}}}$$
\centerline{{\bf TABLE III: Bottom Couplings}}
\smallskip
\noindent {\eightrm The upper half of this table shows the results of fits for
non-standard b-couplings. The global fit, where both $\ss\delta\gl^b$ and
$\ss\delta\gr^b$
are allowed to vary, is shown as well as the two individual fits. Numbers in
brackets
display the pulls. For the global fit the optimal combination of the
b-couplings which
diagonalizes the variance matrix is also given as well as their fit values and
pulls.
In lower half of the table finally it is shown how fit predictions for $\ss
Z^0$-pole
observables change in the global fit. This fit is also shown in figure~\bcoup
.}
%


$$\vbox{\tabskip=0pt \offinterlineskip
\halign to \hsize{\strut#& #\tabskip 1em plus 2em minus .5em&\hfil#\hfil
&\hfil#\hfil &\hfil#\hfil &#\tabskip=0pt\cr
\noalign{\hrule}\noalign{\smallskip}\noalign{\hrule}\noalign{\medskip}
&& \hfil \hbox{Parameter} & \hfil Fit Value \hfil &\hfil Pull \hfil\cr
\noalign{\medskip}\noalign{\hrule}\noalign{\medskip}
&&$\delta\gl^b$&$-0.0011\pm 0.0040$&$0.3$\cr
&&$\delta\gr^b$&$-0.030\pm 0.019$&$1.6$\cr
&&$S$&$-0.25\pm 0.19$&$1.4$\cr
&&$T$&$-0.12\pm 0.21$&$0.6$\cr
\noalign{\medskip}
&&$0.66 S+0.75 T$&$-0.25\pm0.27$&$0.9$\cr
&&$0.98\delta\gl^b -0.18\delta\gr^b$&$-0.0064\pm0.0020$&$3.1$\cr
&&$0.75 S-0.65T-0.13\delta\gr^b$&$-0.114\pm0.074$&$1.5$\cr
&&$0.18\delta\gl^b+0.98\delta\gr^b+0.11S-0.08T$&$0.011\pm 0.017$&$0.7$\cr
\noalign{\medskip}
&& $\as$ & $0.103 \pm 0.008$ &\cr
&& $\mt \hbox{(GeV)}$ & $180$ (fiducial) &\cr
\noalign{\medskip}\noalign{\hrule}\noalign{\medskip}
&& $\chi^2_{\rm min}/\hbox{d.o.f.}$ & 13.7/10 (19\% C.L.) &\cr
\noalign{\medskip}\noalign{\hrule}\noalign{\smallskip}\noalign{\hrule}
}}$$
\centerline{{\bf TABLE IV: Oblique Parameters and Bottom Couplings}}
\smallskip
\noindent {\eightrm This fit shows the results of a oblique fit which also
allows for
new physics contributions to the $\ss Zb\bar{b}$ couplings. As can be seen from
the optimal
linear combinations of the four parameters the two sets, ($\ss S,T$) and ($\ss
\delta
g_{\sss L,R}^b$), are only weakly correlated. The result of this fit, projected
to the
($\ss S,T$)- and ($\ss\delta g_{L,R}^b$)-planes is shown in figure~\obl . Here
a fiducial
value for the top mass has been applied because it cannot be constrained
simultaneously
with the $\ss T$ parameter and non-standard b-couplings.}
%
\vfill\break

$$\vbox{\tabskip=0pt \offinterlineskip
\halign to \hsize{#\tabskip 1em plus 2em minus .5em&
\hfil#\hfil&\hfil#\hfil&\hfil#\hfil&\hfil#\hfil&\tabskip 1em plus 2em minus
.5em #\cr
\noalign{\hrule}\noalign{\smallskip}\noalign{\hrule}\noalign{\medskip}
& \hbox{Parameter} & \hbox{Fit Value}&&\hbox{Pull}&\cr
\noalign{\medskip}\noalign{\hrule}\noalign{\medskip}
&$\delta\gl^b$&$ 0.0003\pm 0.0040 $&& $0.1$&\cr
\noalign{\smallskip}
&$\delta\gr^b$&$ 0.015 \pm 0.019  $&& $0.8$&\cr
\noalign{\smallskip}
&$\delta\gl^c$&$-0.022 \pm 0.010  $&& $2.2$&\cr
\noalign{\smallskip}
&$\delta\gr^c$&$ 0.009 \pm 0.014  $&& $0.7$&\cr
\noalign{\medskip}
&$ 0.13\delta\gl^b+0.94\delta\gr^b+0.02\delta\gl^c-0.32\delta\gr^c$
&$ 0.011\pm 0.020 $&& $0.6$&\cr
\noalign{\smallskip}
&$ 0.97\delta\gl^b-0.17\delta\gr^b+0.17\delta\gl^c-0.07\delta\gr^c$
&$ -0.0067\pm 0.0020 $&& $3.3$&\cr
\noalign{\smallskip}
&$ 0.19\delta\gl^b+0.05\delta\gr^b-0.96\delta\gl^c+0.18\delta\gr^c$
&$ 0.024\pm 0.010 $&& $2.4$&\cr
\noalign{\smallskip}
&$ 0.09\delta\gl^b+0.30\delta\gr^b+0.20\delta\gl^c-0.93\delta\gr^c$
&$ 0.0087\pm 0.0131 $&& $0.7$&\cr
\noalign{\medskip}
&$\alpha_s({\ss M_Z})$&$ 0.180 \pm 0.035 $&&&\cr
\noalign{\smallskip}
&$\mt$ (GeV)&$ 189 \pm 7       $&&&\cr
\noalign{\medskip}
&$\chi^2/{\rm d.o.f.}$&$9.8/9$ ($37\%$ C.L.)&&&\cr
\noalign{\medskip}\noalign{\hrule}\noalign{\medskip}
& Observable & Experimental Value & New Fit Value & New Pull&\cr
\noalign{\medskip}\noalign{\hrule}\noalign{\medskip}
& $\Gamma_\ssz$ (GeV) & $2.4963 \pm 0.0032 $ & $2.4993$ &-0.9&\cr
& $\sigma^h_p $ (nb)  & $41.488 \pm 0.078  $ & $41.462$ & 0.3&\cr
& $R_e$               & $20.797 \pm 0.058  $ & $20.762$ & 0.6&\cr
& $R_\mu$             & $20.796 \pm 0.043  $ & $20.762$ & 0.8&\cr
& $R_\tau$            & $20.813 \pm 0.061  $ & $20.810$ & 0.0&\cr
& $\AFB{e}$           & $0.0157 \pm 0.0028 $ & $0.0165$ &-0.3&\cr
\noalign{\smallskip}
& $\AFB{\mu}$         & $0.0163 \pm 0.0016 $ & $0.0165$ &-0.2&\cr
\noalign{\smallskip}
& $\AFB{\tau}$        & $0.0206 \pm 0.0023 $ & $0.0165$ & 1.8&\cr
\noalign{\smallskip}
& $A_\tau(P_\tau)$    & $0.1418 \pm 0.0075 $ & $0.1484$ &-0.9&\cr
\noalign{\smallskip}
& $A_e(P_\tau)$       & $0.139  \pm 0.0089 $ & $0.1484$ &-1.1&\cr
\noalign{\smallskip}
& $R_b$               & $0.2219 \pm 0.0017 $ & $0.2219$ & 0.0&\cr
& $R_c$               & $0.154  \pm 0.0074 $ & $0.1540$ & 0.0&\cr
& $\AFB{b}$           & $0.0997 \pm 0.0031 $ & $0.0997$ & 0.0&\cr
\noalign{\smallskip}
& $\AFB{c}$           & $0.0729 \pm 0.0058 $ & $0.0729$ & 0.0&\cr
\noalign{\smallskip}
& $\ALR^0$            & $0.1551 \pm 0.0040 $ & $0.1484$ & 1.7&\cr
\noalign{\medskip}\noalign{\hrule}\noalign{\smallskip}\noalign{\hrule}}}$$
\centerline{{\bf TABLE V: Bottom and Charm Couplings}}
\smallskip
\noindent {\eightrm The result of a fit where both $\ss\delta g_{\sss
L,R}^{\sss b}$
and $\ss\delta g_{\sss L,R}^{\sss c}$ are allowed to vary. The lower half of
the table
shows how the fit values of the observables change in this case. Notice the
very high,
although weekly determined, value of the strong coupling constant. Notice also
that
$\ss\delta g_{\sss L}^{\sss c}$ and $\ss\delta g_{\sss R}^{\sss c}$ are only
weekly
correlated, so that the deviation is in the left handed c-coupling only. A fit
similar
to this one, but with $\ss\delta g_R^b$ and $\ss\delta g_R^c$ constrained to
zero is
shown in figure \ccoup .}
%


$$\vbox{\tabskip=0pt \offinterlineskip
\halign to \hsize{\hfil#\hfil&\hfil#\hfil&\hfil#\hfil
&\hfil#\hfil &#\hfil&\hfil#\hfil&\hfil#\hfil\cr
\noalign{\hrule}\noalign{\smallskip}\noalign{\hrule}\noalign{\medskip}
\hfil \hbox{Parameter} & \hfil \hbox{$\alpha_s({\ss M_Z})=0.110$}&&
\hfil \hbox{$\alpha_s({\ss M_Z})$=0.120}&&\hbox{$\alpha_s({\ss
M_Z})=0.130$}&\cr
\noalign{\medskip}\noalign{\hrule}\noalign{\medskip}
$\delta\gl^b$&$-0.0070\pm 0.0017 $& $(4.2)$&$-0.0063\pm 0.0017$& $(3.7)$&
$ -0.0056\pm 0.0017$&$(3.3)$\cr
\noalign{\smallskip}
$\delta\gl^c$&$-0.0187\pm 0.0085 $& $(2.2)$&$-0.0193\pm 0.0085$& $(2.3)$&
$ -0.0200\pm 0.0085$&$(2.4)$\cr
\noalign{\smallskip}
$\delta_s   $&$ 0.0054\pm 0.0028 $& $(1.9)$&$ 0.0045\pm 0.0028$& $(1.6)$&
$ 0.0036 \pm 0.0028$&$(1.3)$\cr
\noalign{\medskip}
$P_1$&$-0.0057 \pm 0.0017 $& $(3.3)$&$ -0.0054 \pm 0.0017 $& $(3.1)$&
$ -0.0050\pm 0.0017$&$(2.9)$\cr
\noalign{\smallskip}
$P_2$&$-0.00154\pm 0.00043$& $(3.6)$&$ -0.00031\pm 0.00043$& $(0.7)$&
$ 0.00091\pm 0.00043$&$(2.1)$\cr
\noalign{\smallskip}
$P_3$&$-0.0198 \pm 0.0089 $& $(2.2)$&$ -0.0201 \pm 0.0089 $& $(2.3)$&
$ -0.0204\pm 0.0089$&$(2.3)$\cr
\noalign{\medskip}
$\chi^2/{\rm d.o.f.}$&$12.7/12$ ($39\%$ C.L.)&&$12.6/12$ ($40\%$ C.L.)&&
12.6/12 ($40\%$ C.L.)&\cr
\noalign{\medskip}\noalign{\hrule}\noalign{\smallskip}\noalign{\hrule}}}$$
\noindent
$P_1\;\equiv\; 0.93\delta\gl^b+0.06\delta\gl^c+0.36\delta\gl^s$\hfil\break
$P_2\;\equiv\; 0.36\delta\gl^b-0.30\delta\gl^c-0.88\delta\gl^s$\hfil\break
$P_3\;\equiv\; 0.06\delta\gl^b+0.95\delta\gl^c-0.30\delta\gl^s$\hfil\break
\bigskip
\centerline{{\bf TABLE VI: Bottom, Charm and Strange Couplings}}
\smallskip
\noindent {\eightrm New physics contributions to $\ss Zs\bar{s}$ can serve to
increase $\ss\Gamma_{\rm had}$ in fits where non standard charm and bottom
couplings
resolve the $\ss R_b$ and $\ss R_c$ discrepancies. The strong coupling constant
$\ss\alpha_s$
is no longer strongly constrained by such fits and can therefore take on values
that are consistent with low-energy determinations. This fit illustrates this
point
by varying $\ss\alpha_s$ in different fits for $\ss\delta\gl^b,\delta\gl^c$ and
$\ss\delta_s$
(defined in equation~\strange ). Here a fiducial value for the top mass, $\ss
m_t=180$ GeV,
has been employed.}
%
\vfill\break

$$\vbox{\tabskip=0pt \offinterlineskip
\halign to \hsize{#\tabskip 1em plus 2em minus .5em&
\hfil#\hfil&\hfil#\hfil&\hfil#\hfil&\hfil#\hfil&\tabskip 1em plus 2em minus
.5em #\cr
\noalign{\hrule}\noalign{\smallskip}\noalign{\hrule}\noalign{\medskip}
& \hbox{Parameter} & \hbox{Fit Value}&&\hbox{Pull}&\cr
\noalign{\medskip}\noalign{\hrule}\noalign{\medskip}
&$\delta u_L$&$ -0.0080\pm 0.0032 $&& $2.5$&\cr
\noalign{\smallskip}
&$\delta u_R$&$ 0.0055 \pm 0.0051  $&& $1.1$&\cr
\noalign{\smallskip}
&$\delta d_L$&$-0.0049 \pm 0.0017  $&& $2.9$&\cr
\noalign{\smallskip}
&$\delta d_R$&$ 0.0115 \pm 0.0059  $&& $1.9$&\cr
\noalign{\medskip}
&$ 0.32\delta u_L+0.56\delta u_R-0.1\delta d_L-0.76\delta d_R$
&$-0.0077\pm 0.0074 $&& $1.0$&\cr
\noalign{\smallskip}
&$ 0.73\delta u_L+0.28\delta u_R+0.41\delta d_L+0.47\delta d_R$
&$-0.0010\pm 0.0022 $&& $0.4$&\cr
\noalign{\smallskip}
&$ 0.46\delta u_L-0.21\delta u_R-0.85\delta d_L+0.15\delta d_R$
&$ 0.00107\pm 0.00033 $&& $3.3$&\cr
\noalign{\smallskip}
&$ 0.39\delta u_L-0.75\delta u_R+0.32\delta d_L-0.43\delta d_R$
&$ -0.0137\pm 0.0038 $&& $3.6$&\cr
\noalign{\medskip}
&$\chi^2/{\rm d.o.f.}$&$15.8/18$ ($61\%$ C.L.)&&&\cr
\noalign{\medskip}\noalign{\hrule}\noalign{\medskip}
& Observable & Experimental Value & New Fit Value & New Pull&\cr
\noalign{\medskip}\noalign{\hrule}\noalign{\medskip}
& $\Gamma_\ssz$ (GeV) & $2.4963 \pm 0.0032 $ & $2.4974$ &-0.4&\cr
& $\sigma^h_p $ (nb)  & $41.488 \pm 0.078  $ & $41.449$ & 0.5&\cr
& $R_e$               & $20.797 \pm 0.058  $ & $20.775$ & 0.4&\cr
& $R_\mu$             & $20.796 \pm 0.043  $ & $20.775$ & 0.5&\cr
& $R_\tau$            & $20.813 \pm 0.061  $ & $20.823$ &-0.2&\cr
& $\AFB{e}$           & $0.0157 \pm 0.0028 $ & $0.0159$ &-0.1&\cr
\noalign{\smallskip}
& $\AFB{\mu}$         & $0.0163 \pm 0.0016 $ & $0.0159$ & 0.3&\cr
\noalign{\smallskip}
& $\AFB{\tau}$        & $0.0206 \pm 0.0023 $ & $0.0159$ & 2.0&\cr
\noalign{\smallskip}
& $A_\tau(P_\tau)$    & $0.1418 \pm 0.0075 $ & $0.1455$ &-0.5&\cr
\noalign{\smallskip}
& $A_e(P_\tau)$       & $0.139  \pm 0.0089 $ & $0.1455$ &-0.7&\cr
\noalign{\smallskip}
& $R_b$               & $0.2219 \pm 0.0017 $ & $0.2218$ & 0.1&\cr
& $R_c$               & $0.154  \pm 0.0074 $ & $0.1632$ &-1.2&\cr
& $\AFB{b}$           & $0.0997 \pm 0.0031 $ & $0.1004$ &-0.2&\cr
\noalign{\smallskip}
& $\AFB{c}$           & $0.0729 \pm 0.0058 $ & $0.0738$ &-0.2&\cr
\noalign{\smallskip}
& $\ALR^0$            & $0.1551 \pm 0.0040 $ & $0.1455$ & 2.4&\cr
\noalign{\smallskip}
& $Q_W$ (Cs)          & $-71.04 \pm  1.81  $ & $ -71.03$ &$0.0$&\cr
& $g^e_A$ ($\nu$-e)   & $-0.503 \pm  0.017 $ & $ -0.507$ &$0.2$&\cr
& $g^e_V$ ($\nu$-e)   & $-0.035 \pm  0.017 $ & $ -0.037$ &$0.1$&\cr
& $C_{1u}-{1\over2}C_{1d}$ (e-D) & $-0.47 \pm  0.13$  & $ -0.355$ &$-0.9$&\cr
\noalign{\smallskip}
& $C_{2u}-{1\over2}C_{2d}$ (e-D) & $0.33 \pm  0.62$  & $ -0.039$ &$0.6$&\cr
& $g_L^2$ ($\nu$ -N)  & $ 0.3003 \pm  0.0039$  & $ 0.3021$ &$-0.5$&\cr
\noalign{\smallskip}
& $g_R^2$ ($\nu$ -N)  & $ 0.0323 \pm  0.0033$  & $ 0.0303$ &$0.6$&\cr
\noalign{\medskip}\noalign{\hrule}\noalign{\smallskip}\noalign{\hrule}}}$$
\centerline{{\bf TABLE VII: Universal Corrections to Up and Down Type 
Quark Couplings}}
\smallskip
\noindent {\eightrm A fit done to see the effect of universal corrections
to up and down type quark couplings as defined in eq.~\universal . Since 
$\ss\alpha_s$ is only very weakly determined in such a fit it has been fixed to
a fiducial value of $\ss\alpha_s = 0.112$. Similarily the top mass has been set
to $\ss m_t = 180 \GeV$. Notice that one needs both, corrections to up and down
type as well as to left- and right-handed couplings to satisfy
constraints set by low-energy observables. Notice also 
that the confidence level
of this fit does not compare directly to those in other fits because here the
low-energy observables, listed in table~I, have also been taken into account.}
%
\vfill\break


$$\vbox{\tabskip=0pt \offinterlineskip
\halign to \hsize{#\tabskip 1em plus 2em minus .5em&\hfil#\hfil&\hfil#\hfil&
\hfil#\hfil&\hfil#\hfil&\hfil#\hfil&\hfil#\hfil&\tabskip 1em plus 2em minus
.5em #\cr
\noalign{\hrule}\noalign{\smallskip}\noalign{\hrule}\noalign{\medskip}
&\hfil \hbox{Parameter} & \hbox{Fit Value}&\hbox{Pull}&
\hbox{Combination}&\hbox{Fit value}&\hbox{Pull}&\cr
\noalign{\medskip}\noalign{\hrule}\noalign{\medskip}
&$\delta\gl^b$    &$-0.0059 \pm 0.0016  $& $3.6$&$P_1$&$ 0.119  \pm 0.095   $&
$1.2$&\cr
\noalign{\smallskip}
&$\delta\gl^e$    &$-0.00068\pm 0.00067 $& $1.0$&$P_2$&$ -0.0010 \pm 0.0079  $&
$0.1$&\cr
\noalign{\smallskip}
&$\delta\gr^e$    &$-0.00059\pm 0.00057 $& $1.0$&$P_3$&$ 0.0009 \pm 0.0024  $&
$0.4$&\cr
\noalign{\smallskip}
&$\delta\gl^u$    &$ 0.065  \pm 0.051   $& $1.3$&$P_4$&$ 0.0046 \pm 0.0020  $&
$2.3$&\cr
\noalign{\smallskip}
&$\delta\gr^u$    &$-0.041  \pm 0.034   $& $1.2$&$P_5$&$-0.0040 \pm 0.0017  $&
$2.3$&\cr
\noalign{\smallskip}
&$\delta\gl^d$    &$ 0.056  \pm 0.043   $& $1.3$&$P_6$&$ 0.00169\pm 0.00066 $&
$2.6$&\cr
\noalign{\smallskip}
&$\delta\gr^d$    &$-0.073  \pm 0.060   $& $1.2$&$P_7$&$ 0.00039\pm 0.00027 $&
$1.4$&\cr
\noalign{\smallskip}
&$\Delta^{\rm LE}$&$ 0.00065\pm 0.00133 $& $0.5$&$P_8$&$-0.00074\pm 0.00041 $&
$1.8$&\cr
\noalign{\medskip}
&$\alpha_s$    &$ 0.112$&&(fiducial)&&&\cr
&$m_t$    &$ 180$&&(fiducial)&&&\cr
\noalign{\medskip}
&$\chi^2/{\rm d.o.f.}$&$16.6/14$ ($28\%$ C.L.)&&\cr
\noalign{\medskip}\noalign{\hrule}\noalign{\smallskip}\noalign{\hrule}}}$$
\noindent
$P_1\;\equiv\;  0.54\delta\gl^u -0.30\delta\gr^u +0.45\delta\gl^d
-0.62\delta\gr^d $\hfil\break
$P_2\;\equiv\;  0.12\delta\gl^u +0.78\delta\gr^u -0.29\delta\gl^d
-0.54\delta\gr^d$\hfil\break
$P_3\;\equiv\;0.18\delta\gl^b-0.06\delta\gl^e-0.03\delta\gr^e+0.71
                 \delta\gl^u+0.34\delta\gr^u
                +0.18\delta\gl^d +0.55\delta\gr^d +0.13\Delta^{\rm LE}
$\hfil\break
$P_4\;\equiv\;-0.58\delta\gl^b+0.11\delta\gl^e+0.07\delta\gr^e-0.20
                 \delta\gl^u+0.35\delta\gr^u
                +0.65\delta\gl^d +0.10\delta\gr^d -0.24\Delta^{\rm LE}
$\hfil\break
$P_5\;\equiv\;0.60\delta\gl^b+0.31\delta\gl^e+0.21\delta\gr^e-0.08
                 \delta\gl^u+0.07\delta\gr^u
                +0.16\delta\gl^d -0.68\Delta^{\rm LE} $\hfil\break
$P_6\;\equiv\;-0.50\delta\gl^b+0.15\delta\gl^e+0.41\delta\gr^e+0.37
                 \delta\gl^u-0.18\delta\gr^u
                -0.47\delta\gl^d +0.08\delta\gr^d -0.41\Delta^{\rm LE}
$\hfil\break
$P_7\;\equiv\;  -0.11\delta\gl^b
+0.74\delta\gl^e-0.65\delta\gr^e+0.09\delta\gl^u-0.04\delta\gr^u
                -0.11\delta\gl^d +0.02\delta\gr^d$\hfil\break
$P_8\;\equiv\; 0.08\delta\gl^b +0.56\delta\gl^e
+0.61\delta\gr^e-0.07\delta\gl^u+0.03\delta\gr^u
                +0.08\delta\gl^d -0.01\delta\gr^d +0.55\Delta^{\rm LE}
$\hfil\break
\bigskip
\centerline{{\bf TABLE VIII: First Generation Couplings}}
\smallskip
\noindent {\eightrm In order to constrain non-standard contributions to first
generation
couplings, the observables from low energy neutral current experiments shown in
the second half
of table~I have been included in this fit. Because of this the confidence level
of this fit
compares not directly with the ones in other tables. Also both, the top mass
and $\ss\alpha_s$
have been fixed to fiducial values. Notice that the three optimal combinations
that deviate
the most, $\ss P_4$ to $\ss P_6$, all receive large contributions from
$\ss\delta g_L^b$.
}
%
\vfill\break


$$\vbox{\tabskip=0pt \offinterlineskip
\halign to \hsize{#\tabskip 1em plus 2em minus .5em&\hfil#\hfil&\hfil#\hfil&
\hfil#\hfil&\hfil#\hfil&\hfil#\hfil&\hfil#\hfil&\tabskip 1em plus 2em minus
.5em #\cr
\noalign{\hrule}\noalign{\smallskip}\noalign{\hrule}\noalign{\medskip}
&\hfil \hbox{Parameter} & \hbox{Fit
Value}&\hbox{Pull}&\hbox{Combination}&\hbox{Fit value}
&\hbox{Pull}&\cr
\noalign{\medskip}\noalign{\hrule}\noalign{\medskip}
&$\delta\gl^b$&$-0.0070 \pm 0.0017  $& $4.1$&$P_1$&$-0.003  \pm 0.013   $&
$0.2$&\cr
\noalign{\smallskip}
&$\delta\gl^c$&$-0.0171\pm 0.0089  $& $1.9$&$P_2$&$ 0.0202 \pm 0.0087  $&
$2.3$&\cr
\noalign{\smallskip}
&$\delta\gr^c$&$ 0.009\pm 0.013 $& $0.7$&$P_3$&$ 0.0056 \pm 0.0017  $&
$3.3$&\cr
\noalign{\smallskip}
&$\delta\gl^\mu$&$ 0.0000  \pm 0.0017   $& $0.0$&$P_4$&$-0.00072 \pm 0.00055
$& $1.3$&\cr
\noalign{\smallskip}
&$\delta\gr^\mu$&$-0.0004  \pm 0.0020   $& $0.2$&$P_5$&$ 0.00092 \pm 0.00033
$& $2.8$&\cr
\noalign{\smallskip}
&$\delta_s$&$ 0.0057  \pm 0.0030   $& $1.9$&$P_6$&$-0.0003\pm 0.0025 $&
$2.5$&\cr
\noalign{\medskip}
&$\alpha_s$    &$ 0.112$&&(fiducial)&&&\cr
&$m_t$    &$ 180$&&(fiducial)&&&\cr
\noalign{\medskip}
&$\chi^2/{\rm d.o.f.}$&$11.9/9$ ($22\%$ C.L.)&&\cr
\noalign{\medskip}\noalign{\hrule}\noalign{\smallskip}\noalign{\hrule}}}$$
\noindent
$P_1\;\equiv\;0.01\delta\gl^b-0.33\delta\gl^c-0.94\delta\gr^c
              -0.03\delta_s $\hfil\break
$P_2\;\equiv\;-0.06\delta\gl^b-0.89\delta\gl^c+0.31\delta\gr^c
              +0.33\delta_s $\hfil\break
$P_3\;\equiv\;-0.94\delta\gl^b-0.05\delta\gl^c+0.02\delta\gr^c
              -0.35\delta_s $\hfil\break
$P_4\;\equiv\;0.27\delta\gl^b-0.24\delta\gl^c+0.11\delta\gr^c
              +0.47\delta\gl^\mu-0.40\delta\gr^\mu-0.70\delta_s $\hfil\break
$P_5\;\equiv\;-0.22\delta\gl^b+0.19\delta\gl^c-0.08\delta\gr^c
              +0.60\delta\gl^\mu-0.51\delta\gr^\mu+0.54\delta_s $\hfil\break
$P_6\;\equiv\;0.65\delta\gl^\mu+0.76\delta\gr^\mu$\hfil\break
\bigskip
\centerline{{\bf TABLE IX: Second Generation Couplings}}
\smallskip
\noindent {\eightrm This fit is not very different from the one shown in
table~VI.
In addition to the couplings floated there it also allows for non-standard
$\ss Z^0\mu\bar{\mu}$ couplings. As can be seen above these agree almost
perfectly
with zero.
}
%
\vfill\break


$$\vbox{\tabskip=0pt \offinterlineskip
\halign to \hsize{#\tabskip 1em plus 2em minus .5em&\hfil#\hfil&\hfil#\hfil&
\hfil#\hfil&\tabskip 1em plus 2em minus .5em #\cr
\noalign{\hrule}\noalign{\smallskip}\noalign{\hrule}\noalign{\medskip}
&\hfil \hbox{Parameter} & \hbox{Fit Value}&\hbox{Pull}&\cr
\noalign{\medskip}\noalign{\hrule}\noalign{\medskip}
&$\delta\gl^b$    &$-0.0029 \pm 0.0037  $& $0.8$&\cr
\noalign{\smallskip}
&$\delta\gr^b$    &$0.023\pm 0.019 $& $1.2$&\cr
\noalign{\smallskip}
&$\delta\gl^\tau$    &$ 0.00018  \pm 0.00100   $& $0.2$&\cr
\noalign{\smallskip}
&$\delta\gr^\tau$    &$ 0.00017  \pm 0.00106   $& $0.2$&\cr
\noalign{\medskip}
&$0.166\delta\gl^b+0.986\delta\gr^b$
&$ 0.022  \pm 0.019   $& $1.2$&\cr
\noalign{\smallskip}
&$0.984\delta\gl^b-0.165\delta\gr^b-0.047\delta\gl^\tau-0.041\delta\gr^\tau$
&$ -0.0066  \pm 0.0021   $& $3.2$&\cr
\noalign{\smallskip}
&$-0.06\delta\gl^b+0.02\delta\gr^b-0.67\delta\gl^\tau-0.74\delta\gr^\tau$
&$ 0.0004  \pm 0.0013   $& $0.3$&\cr
\noalign{\smallskip}
&$0.74\delta\gl^\tau-0.67\delta\gr^\tau$
&$-0.00009  \pm 0.00056   $& $0.2$&\cr
\noalign{\medskip}
&$\alpha_s$    &$ 0.101  \pm 0.008   $&&\cr
&$m_t$    &$ 189  \pm 7   $&&\cr
\noalign{\medskip}
&$\chi^2/{\rm d.o.f.}$&$15.4/9$ ($8\%$ C.L.)&&\cr
\noalign{\medskip}\noalign{\hrule}\noalign{\smallskip}\noalign{\hrule}}}$$
\noindent
\bigskip
\centerline{{\bf TABLE X: Third Generation Couplings}}
\smallskip
\noindent {\eightrm A fit to the third generation couplings. Notice that the
$\ss Z\tau\bar{\tau}$, in agreeing very well with zero, do not help much to
resolve
any of the discrepancies observed in $\ss A_{FB}^0(\tau )$ and other
observables,
resulting in a low confidence level.
}
%

\figurecaptions

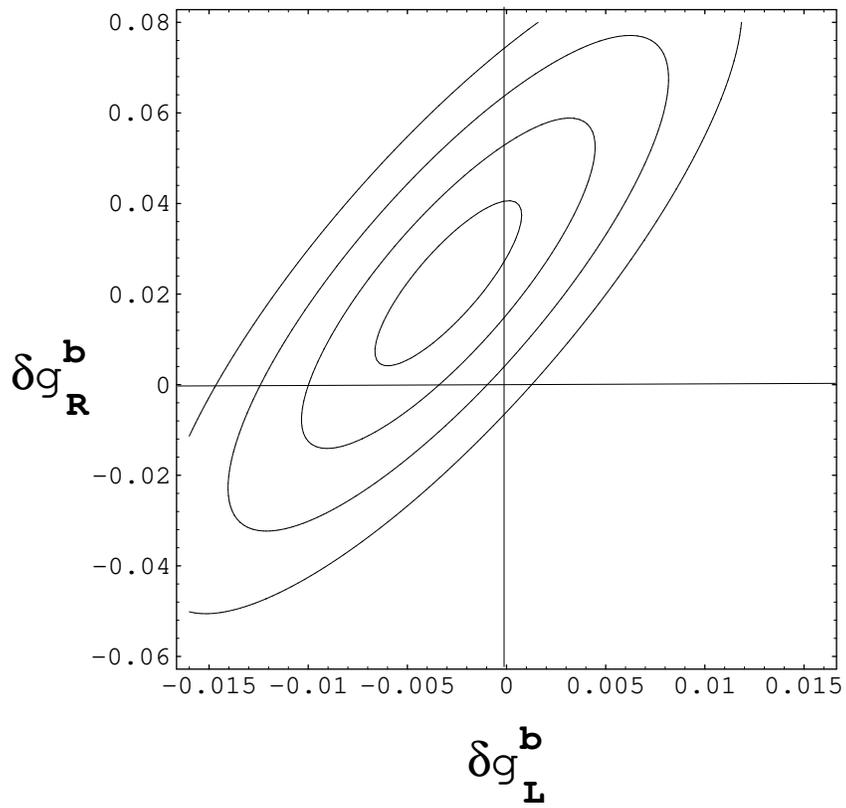
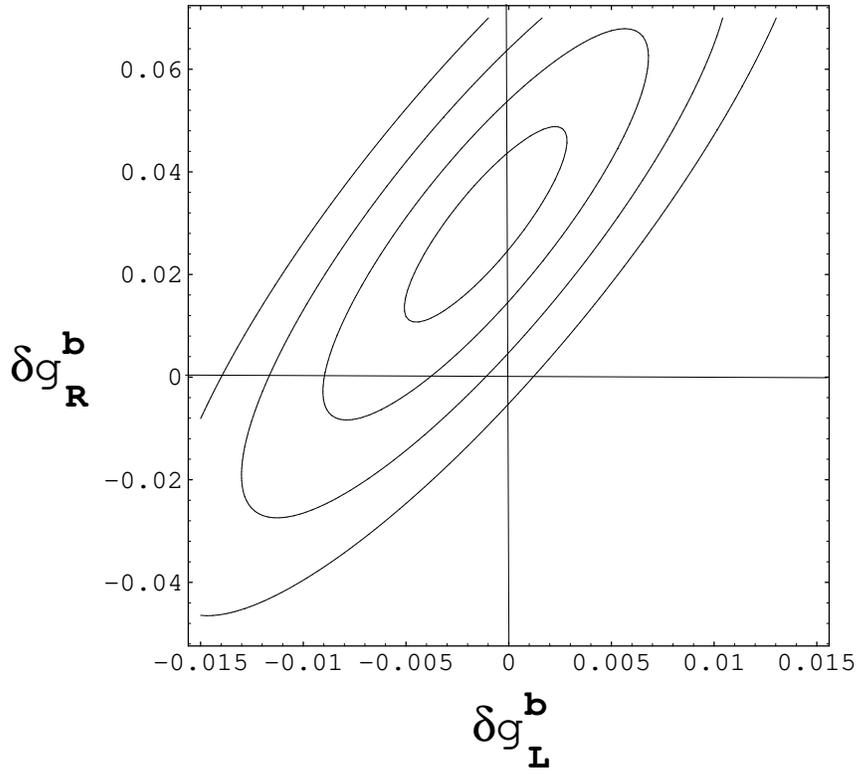
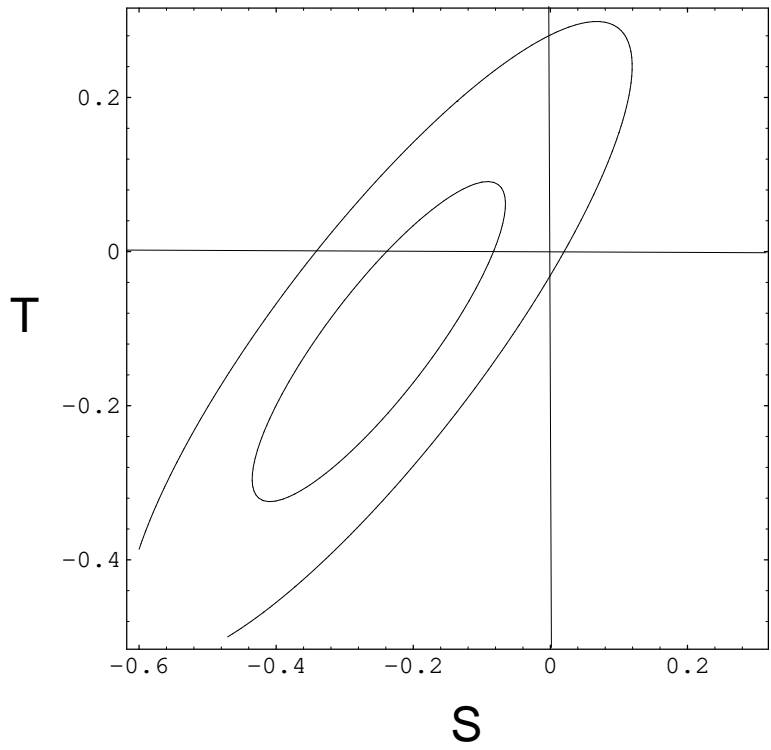
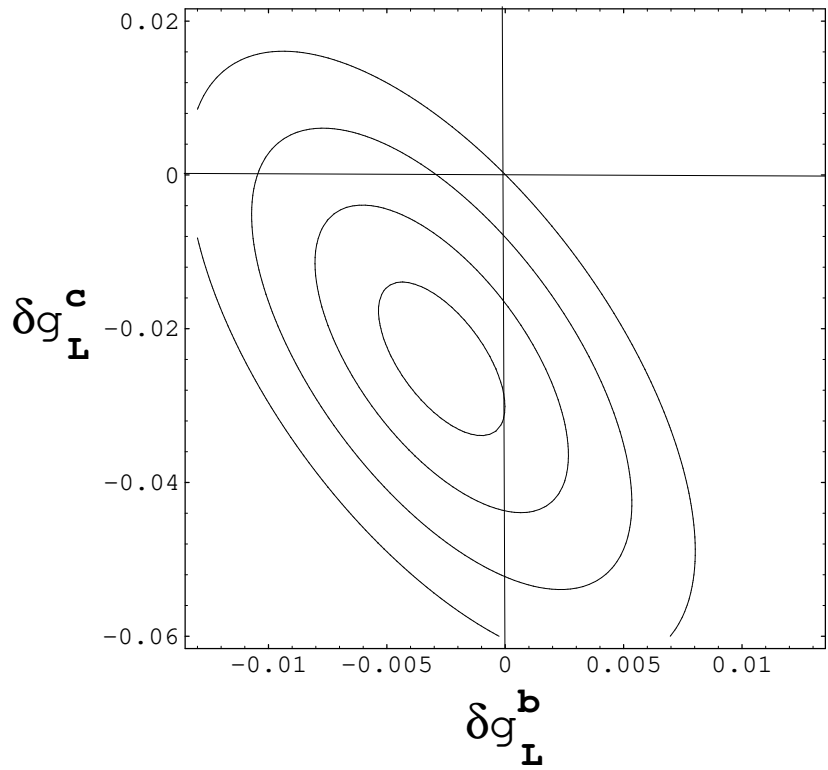
\bye